\documentclass[aps,prl,superscriptaddress,floatfix,amsmath,amsfonts,twocolumn]{revtex4-2}  

\usepackage{amsfonts}
\usepackage{upgreek}
\usepackage{braket}
\usepackage{verbatim}
\usepackage{xr}
\usepackage{enumitem}
\usepackage{amsmath}
\usepackage{graphicx}
\usepackage{epsfig}
\usepackage{bm}
\usepackage{bbold}
\usepackage{amssymb}
\usepackage{color}
\usepackage[normalem]{ulem}  %

\newcommand{\vep}{\varepsilon}

\newcommand{\micro}{${\upmu}$}

\newcommand{\um}{$\,$\micro m}

\begin{document}

\title{Transport regimes for exciton-polaritons in disordered microcavities} 

\author{A.N. Osipov}
\affiliation{Department of Physics, ITMO University, Saint Petersburg 197101, Russia}

\author{I.V. Iorsh}
\affiliation{Department of Physics, ITMO University, Saint Petersburg 197101, Russia}
\affiliation{Abrikosov Center for Theoretical Physics, MIPT, Dolgoprudnyi, Moscow Region 141701, Russia}

\author{A.V. Yulin}
\affiliation{Department of Physics, ITMO University, Saint Petersburg 197101, Russia}
\affiliation{Science Institute, University of Iceland, Dunhagi 3, IS-107, Reykjavik, Iceland}

\author{I.A. Shelykh}
\affiliation{Science Institute, University of Iceland, Dunhagi 3, IS-107, Reykjavik, Iceland}
\affiliation{Department of Physics, ITMO University, Saint Petersburg 197101, Russia}
\affiliation{Abrikosov Center for Theoretical Physics, MIPT, Dolgoprudnyi, Moscow Region 141701, Russia}

\date{\today}

\begin{abstract}
%
Light-matter coupling in a planar optical cavity substantially  modifies  the transport regimes in the system in presence of a short range excitonic disorder. Basing on Master equation for a resonantly coupled exciton-photon system, and treating disorder scattering in the Born-Markov approximation we demonstrate the onset of ballistic and diffusive transport regimes in the limits of weak and strong disorder respectively. We show that transport parameters governing the crossover between these two regimes strongly depend on the parameters characterizing light-matter coupling, in particular Rabi energy and detuning between excitonic and photonic modes. The presented theory agrees with recent experimental data on transport in disordered organic microcavities. 
\end{abstract}

\maketitle

\section{Introduction}
Bright excitons are bound states of electrons and holes, which can be created optically in direct bandgap materials. These composite quasiparticles can be used for a variety of optoelectronic applications \cite{Sanvitto2001, Gregg2003, Banappanavar2021}. There exist, however, certain obstacles, which make direct applications of excitons problematic. In particular, being massive particles optically created excitons usually have small group velocity. In realistic samples, where disorder is always present, this leads to very short dephasing times and, consequently short characteristic propagation lengths of excitons, which becomes a major problem in such fields as photovoltaics \cite{Classen2020, Gillet2021}. 

Excitonic transport and the ways to enhance it were widely studied in literature \cite{akselrod2014visualization, mikhnenko2015exciton, kulig2018exciton, fortin1993exciton, deng2020long, rudolph2007long}. An attractive option consists in resonant coupling of an excitonic transition to a photonic mode of an optical cavity. If the energy of this coupling exceeds all characteristic broadenings in the system, the regime of strong light-matter coupling is established and novel type of quasiparticles, exciton-polaritons, are formed. Being of hybrid nature, polaritons combine the properties of light and matter particles forming them. In particular, from the photonic component they inherit extremely small effective mass (about $10^{-5}$ of the mass of free electrons) and macroscopically large coherence length \cite{Ballarini2017}, while
the presence of the excitonic component enables the sensitivity of polaritons to external potentials and, in particular, allows efficient polariton scattering on excitonic disorder potential \cite{Borri2000}.

 Recently, it has been observed that mixing with the photonic mode significantly modifies the transport properties of polaritons at low temperatures as compared to bare excitons in both organic \cite{balasubrahmaniyam2023enhanced, myers2018polariton, orgiu2015conductivity, lerario2017high, hou2020ultralong, rozenman2018long}, and inorganic \cite{wurdack2021motional, guo2022boosting} microcavities.   In general, in samples with high photonoic fractions ballistic transport regime is usually established, whereas crossover to the regime of diffusive transport occurs when excitonic fraction is increased, as it was recently unambiguously demonstrated in Ref. \onlinecite{balasubrahmaniyam2023enhanced}.  
 
 This effect has clear qualitative explanation. Indeed, at low temperature when short range impurity scattering gives major contribution to the transport, it affects only the excitonic part of a wavefunction of a polariton, while photonic part remains coherent. This results in experimentally observable narrowing of a polariton's linewidth due to the suppression of excitonic inhomogeneous broadening \cite{wurdack2021motional}, the effect known as polariton motional narrowing. While first theoretical description of this effect have been developed decades ago~\cite{PhysRevLett.78.4470}, a microscopic theory of the crossover between different transport regimes in the real space related to it is still lacking. Creation of such a theory is an actual task, specifically in light of the recent revival of  experimental activity in this field ~\cite{balasubrahmaniyam2023enhanced, wurdack2021motional}.

In the present work we aim to fill this gap. We use the density matrix formalism \cite{PhysRevB.83.165316, carmichael2013statistical, grimaldi2005electron, averkiev2008specific},  to derive master equation for polaritons in microcavities in the presence of randomly located impurities. We show that for excitons equations describing both diffusive \cite{kulig2018exciton, glazov2019phonon}, and ballistic propagation \cite{agranovich2007nature} can be derived from the master equation in the limits strong and weak disorder respectively. For polaritons, we use adiabatic elimination technique \cite{brion2007adiabatic} to get rid of the upper polariton branch in the limit when characteristic energy of the disorder potential is smaller then the Rabi splitting. The analysis of the dynamics of lower polaritons allowed us to demonstrate that the increase of the photonic fraction enhances group velocity of the excitations and suppresses the scattering on a short range disorder potential, which leads to a crossover from the diffusive to ballistic propagation regime. Expressions for relevant quantitative characteristics describing such a crossover, such as polaritonic relaxation time and diffusion coefficient \cite{lifschitz1983physical, altland2010condensed} are derived. Our results contribute to the understanding of the relation between motional narrowing and regimes of the polariton dynamics in the real space. 



\section{The model} 
\begin{figure}[tb!]
\begin{center}
\includegraphics[width=\linewidth]{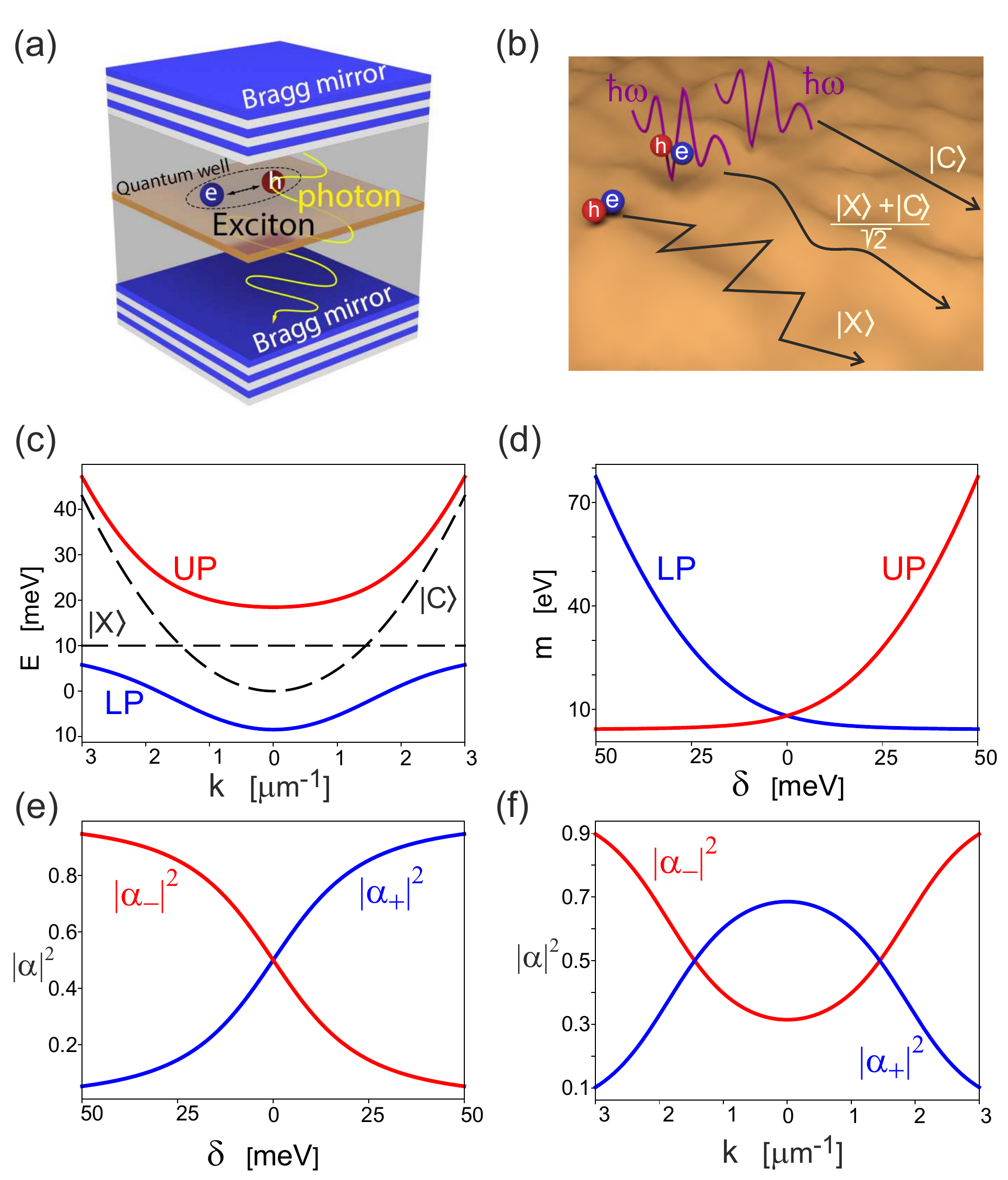}
\end{center}
\caption{ \label{illustration}
(a) Schematic illustration of a microcavity consisting of two Bragg mirrors and a quantum well with an excitonic transition brought close to the resonance with a confined photonic mode.
(b) The illustration of propagation of different types of the excitations in the system in presence of a random  short range potential. Bare excitons have small group velocity, experience strong scattering on the disordered potential and, therefore, are subject to random walks leading to the onset of the diffusive transport regime. On the contrary, photons have very high group velocities, do not experience any scattering on short range potential at all and propagate ballistically. Polaritons are hybrid particles for which an intermediate transport regime is established, which becomes closer to ballistic or diffusive depending on excitonic and photonic fractions which can be controlled by change of a cavity detuning. 
(d) Dispersion characteristics of upper (red solid line) and lower (blue solid line) polaritons calculated for the following parameters  of the system: photonic effective mass $m_{ph} = 0.8\cdot10^{-5}m_e$, Rabi energy \; $\hbar\Omega_R = $ 12.5  meV,  cavity detuning $\delta = $ -10 meV. The dashed black lines show the dispersion of the non-interacting photons and excitons. The hybridization is strongest at the wavevectors corresponding to the crossing of the photonic and excitonic dispersions. Changing the detuning between the resonant frequency of the cavity and the exciton frequency $\Delta E$ it is possible to control both the effective mass of the polaritons (panel (e)) and excitonic and photonic fractions (Hopfield coefficients, panel (g)).  Note, that for a given value of the detuning Hopfield coefficients depend on in-plane momentum of a polariton $k$ (panel (f)).
}
\end{figure}

We consider a 2D planar microcavity formed by two Bragg mirrors with a quantum well (QW) with an excitonic transition embedded in an antinode of a confined cavity mode and brought close to the resonance with it, as it is shown schematically in Fig.\ref{illustration}. We neglect the effects of polariton nonlinearities in the present study, a finite lifetime is not expected to modify the transport regimes and is neglected in our further discussion. Moreover, the lifetime of  photons in in high quality samples can be as long as hundreds of picoseconds \cite{Sun2017}. 

The resonant interaction between excitons and photons leads to the establishment of the strong coupling regime and formation of cavity polaritons, which will be in focus of our attention. Initially coherent polariton wavepacket with controllable parameters can be created in the system by a focused pulse of a coherent light. The presence of a short range disorder in the QW will create an effective random scattering potential affecting the excitonic part of the polariton wavefunction, and polaritons will thus gradually lose their coherence. This will affect their real space dynamics, which will change from ballistic to diffusive, as we will show below. 

In the linear regime, when exciton-exciton interactions can be neglected, the dynamics of the system can be described with the following model Hamiltonian:
\begin{eqnarray}
    &H = \sum_k\Big[\vep_x(\bm{k})b_{\bm{k}}^\dagger b_{\bm{k}} + \vep_c(\bm{k})a_{\bm{k}}^\dagger a_{\bm{k}} +\nonumber \\ 
    &+\hbar\Omega_R (b_{\bm{k}}^\dagger a_{\bm{k}} + a_{\bm{k}}^\dagger b_{\bm{k}})\Big] + \sum_{\bm{kk'}}V_{\bm{kk'}}b_{\bm{k'}}^\dagger b_{\bm{k}},
\end{eqnarray}
where $b_{k},\, a_{k}$ are excitonic and photonic field operators, $\vep_x(\bm{k}),\, \vep_c(\bm{k})$ - the dispersions of bare excitons and photons, $\Omega_R$ is the Rabi frequency controlling the strength of exciton-photon interaction and, finally, $V_{kk'}$ is the matrix element of a short range excitonic disorder potential. In our further consideration we approximate the photonic dispersion by a parabola, $\vep_c(k)=\hbar^2k^2/2m_{ph}$, where $m_{ph}$ is an effective mass of a cavity photon, and take the excitonic dispersion flat, $\vep_x(\bm{k})=\delta=const$. The distance 
\begin{equation}
\delta=\vep_c(0)-\vep_x(0)  
\end{equation}
is an important parameter of the system governing the percentage of excitonic and photonic fractions in a polariton. 

For our purposes it is convenient to represent the Hamiltonian as a sum of the unperturbed Hamiltonian $H_0$ describing excitons and photons in a spatially uniform system and the perturbation $H_V$ accounting for the interaction of excitons with the disorder,
\begin{equation}
    H = H_0 + H_V,
\end{equation}
where 
\begin{equation}
    H_0 = \sum_k\left( \vep_x(\bm{k})b_{\bm{k}}^\dagger b_{\bm{k}} + \vep_c(\bm{k})a_{\bm{k}}^\dagger a_{\bm{k}} + \hbar\Omega_R (b_{\bm{k}}^\dagger a_{\bm{k}} + a_{\bm{k}}^\dagger b_{\bm{k}})  \right),
\end{equation}
and
\begin{equation}
   H_V = \sum_{\bm{kk'}}V_{\bm{kk'}}b_{\bm{k'}}^\dagger b_{\bm{k}}.
\end{equation}

The first part of the Hamiltonian $H_0$ can be diagonalized by moving to the polaritonic basis with use of the unitary transformation
\begin{subequations}
    \begin{eqnarray}
        &&{c_{\bm{k}}}_+ = \alpha_+ b_{\bm{k}} + \alpha_- a_{\bm{k}}, \, \\
        &&{c_{\bm{k}}}_- = \alpha_+a_{\bm{k}} - \alpha_- b_{\bm{k}},
    \end{eqnarray}
\end{subequations}
where ${c_{\bm{k}}}_\pm$ are operators of upper and lower polaritons, $\alpha_\pm$ are Hopfield coefficients corresponding to excitonic and photonic fractions in them. 

As the result one gets
\begin{equation}
    H_0 = \sum_{\bm{k}}\left(E_+{c_{\bm{k}}^\dagger}_+{c_{\bm{k}}}_+ + E_-{c_{\bm{k}}^\dagger}_-{c_{\bm{k}}}_-\right),
\end{equation}
where
\begin{equation}
    E_{\pm}(\bm{k}) = \frac{\vep_c(\bm{k})-\vep_x(\bm{k})}{2}\pm\sqrt{\left(\frac{\vep_c(\bm{k})-\vep_x(\bm{k})}{2}\right)^2 + (\hbar\Omega_R)^2},
\end{equation}
are dispersions of the polariton modes. $E_\pm,\alpha_\pm$ are illustrated by panels (c)-(f) of Fig. 1.

To describe the dynamics in our system, we start from the Liouville-von Neumann equation for the full density matrix,
\begin{equation}
    \partial_t\rho = -\frac{i}{\hbar}[H_0, \rho(t)]  - \frac{i}{\hbar}[H_V, \rho(t)],
\end{equation}
which we treat in Born-Markov approximation \cite{PhysRevB.83.165316}, \cite{carmichael2013statistical}. 
This allows us to get the following master equation:
\begin{eqnarray}
   &&\partial_t\rho = -\frac{i}{\hbar}[H_0, \rho(t)] -\\
   &&-\langle M_0\frac{1}{\hbar^2}[H^I_V(t), \int^t_0 dt'[H^I_V(t'), \rho^I(t)]]M_0^{\dagger}\rangle_c \nonumber, 
   \label{general Master equation}
\end{eqnarray}
where 
\begin{equation}
M_0 = \exp(-\frac{i}{\hbar} H_0 t)    
\end{equation} 
and 
\begin{equation}
H^I(t) = M_0^{\dagger} H^I_V(t)M_0    
\end{equation} 
denotes the scattering Hamiltonian in the interaction picture. The brackets $\langle\rangle_c$ denotes averaging on the non-correlated impurity's position (for details see \textbf{Appendix A}).

Spatio-temporal dynamics of a polariton ensemble is determined by a time evolution of a single particle polariton density matrix \cite{PhysRevB.83.165316}
\begin{equation}
\rho_{\zeta_1, \zeta_2}(\bm{r},\bm{r}',t)= \frac{(2\pi)^2}{A}\int\rho_{\zeta_1, \zeta_2}(\bm{k},\bm{k'},t)e^{i(\bm{k'}\bm{r'}-\bm{k}\bm{r})},
\end{equation}
where $\zeta_{1,2}=\pm$ corresponds to the upper and lower polariton branches, $A$ is an area of a sample and
\begin{equation}
\rho_{\zeta_1, \zeta_2}(\bm{k}, \bm{k'}, t) = \langle{c^\dagger_{\bm{k}}}_{\zeta_1} {c_{\bm{k'}}}_{\zeta_2}\rangle = \text{Tr}(\rho {c^{\dagger}_{\bm{k}}}_{\zeta_1} {c_{\bm{k'}}}_{\zeta_2}),    
\label{Correlators}
\end{equation}

Note, that $\rho_{-, -}(\bm{r},\bm{r},t)$ and $\rho_{+, +}(\bm{r},\bm{r},t)$ correspond to the densities of lower and upper polaritons in the real space, while  $\rho_{-, -}(\bm{k},\bm{k},t)$ and $\rho_{+, +}(\bm{k},\bm{k},t)$ to corresponding occupancies in the k-space. The terms $\rho_{+, -}(\bm{k},\bm{k},t)$ and $\rho_{-, +}(\bm{k},\bm{k},t)$ describe the inter-branch correlations.

The dynamic equations (\ref{general Master equation}) for the correlators defined in Eq.(\ref{Correlators}) read:
\begin{eqnarray}
&&\partial_{t}\rho_{\zeta_1, \zeta_2}(\bm{k}, \bm{k'}, t) =  \frac{i}{\hbar} (E_{\zeta_1}(\bm{k}) - E_{\zeta_2}(\bm{k'}))\rho_{\zeta_1, \zeta_2}(\bm{k}, \bm{k'}, t) - \nonumber \\
&&- \frac{1}{\hbar^2} S,
\label{Final Polaritons's Master equation}
\end{eqnarray}
%
Without the last term in the right-hand side this equation describes coherent (ballistic) polariton propagation. The diffusion of the polaritons occurs due to the scattering on the impurities and this effect is accounted by the second term in Eq.(\ref{Final Polaritons's Master equation}). 
The detailed derivation of (\ref{Final Polaritons's Master equation}) and the expression for the scattering term $S$ can be found in \textbf{Appendix B}). It worse to notice, that the first part of the equation (\ref{Final Polaritons's Master equation}) without scattering term $S$ is analog of the Shrodinger equation and, therefore, equation (\ref{Final Polaritons's Master equation}) can be applied not only for description of transport effects but also for description of quantum effects.

Let us make a remark on the applicability of the perturbation theory. The particle transport can be described within the Born-Markov approximation provided that $\frac{<E>\tau_0}{\hbar} >> 1$ \cite{lifschitz1983physical, averkiev2008specific} where $<E>$ denotes average particles energy, $\tau_0$ is the relaxation time defined in (\ref{relaxation time definition}). A similar approach based on the derivation of the equations for 2x2 density matrix was implemented in \cite{grimaldi2005electron, averkiev2008specific} for the problem of polariton's spin dynamics.

\section{ Exciton transport}
Before we analyze in detail the case of polaritons, where the role of photonic fraction is essential, let us consider the simpler case of bare excitons. We set the Rabi frequency  to zero , $\Omega_R=0$ and thus reduce the problem to the evolution of a single scalar bosonic field.  

The dynamic equations (\ref{Final Polaritons's Master equation}) then reduce to 
\begin{equation}
    \frac{\partial}{\partial t}\rho(\bm{k}, \bm{k'}, t) =  \frac{i}{\hbar} (E(\bm{k}) - E(\bm{k'}))\rho(\bm{k}, \bm{k'}, t) - \frac{1}{\hbar^2}S,
    \label{Excitonic Master equation}
\end{equation}
where
%
%
%
\begin{eqnarray}
    &&S = \pi  \frac{n}{A}\hbar \sum_{\bm{q}}|U_{\bm{q}}|^2((\rho(\bm{k}, \bm{k'}, t)-\rho(\bm{k+q}, \bm{k'+q}, t))\times \notag\\
    &&\times(\delta(E(\bm{k})-E(\bm{k+q}))+ \delta(E(\bm{k'})-E(\bm{k'+q}))).
    \label{Excitonic scattering integral in elastic limit}
\end{eqnarray}
In these formulae $n$ is the impurities concentration that appears in the scattering term after averaging with respect to random impurity's positions, $U_{\bm{q}} =  \int d^2\bm{r} e^{-i\bm{q}\bm{r}}V(\bm{r})$ is a single impurity potential's Fourier component and $V(r)$ - single impurity's potential. The quantity $A$ is the sample area defining the allowed values of $q$. Taking the limit $A \rightarrow \infty$ the summation over $q$ can be substituted with integration  $\sum_{\bm{q}}\rightarrow \frac{A}{(2\pi)^2}\int d^2\bm{q}$. So the area $A$ cancels out from the expression for the  scattering rate which becomes proportional to ($n|U_q|^2$).   


 Equation (\ref{Excitonic Master equation}) with the scattering term (\ref{Excitonic scattering integral in elastic limit}) well describe two transport regimes in the limits of weak and strong disorder.

The first one corresponds to the ballistic transport. Indeed, for a spatially uniform system $U_q = 0$ and then the equation (\ref{Excitonic Master equation}) is nothing else but a well known Schrodinger equation written for the density matrix of a pure state in k-representation.
In this regime one recovers a standard dispersion of a wavepacket corresponding to a massive quantum particle. The size of an envelope $\Delta r(t)$ given by average radius for axially symmetric distributions with $\bm{k}_0 =$ 0 scales linearly with time \cite{agranovich2007nature}. 
\begin{equation}
\Delta r (t) = <r> \xrightarrow{t \rightarrow \infty} a t.   \label{DeltaBallistic} 
\end{equation} 


In the second limit of strong disordered potential, dephasing which accompanies the impurity scattering leads to the fast suppression of the correlations between states corresponding to different $\bm{k}$, so that nonzero elements of the matrix of the correlators $\rho(\bm{k},\bm{k}')$ group around its diagonal (\textbf{See Fig \ref{k_r distributions} in the Appendix}). The transport is now described by kinetic equations of the Boltzmann type, which can be derived from Eq.(\ref{Excitonic Master equation}) by moving to Wigner representation \cite{altland2010condensed}. The transport of excitons is purely diffusive \cite{glazov2019phonon} with asymptotic for the beam width scaling as
\begin{equation}
\Delta r (t) \xrightarrow{t \rightarrow \infty} a t^{0.5}.    \label{DeltaDiffusive}
\end{equation}

To check that the asymptotics above are correct and to consider the intermediate case of a mixed transport, we performed the numerical simulations of two-dimensional excitonic propagation for different disorder strengths characterized by the parameter $n|U|^2$. We took excitonic mass to be twice the mass of a free electron and used s-wave approximation for the disorder scattering, taking the corresponding matrix element to be $q$-independent, $U_\mathbf{q}=U$. Initial distribution of excitons was taken in the form of a coherent Gaussian packet 
\begin{equation}
\rho(t = 0) \sim e^{-\frac{k_r^2+{k_r'}^2}{2\delta k_r}}, \label{InitialRho}   
\end{equation} 
with $\delta k_r =$ 0.085 \um{}$^{-1}$. For details about numerical procedure see \textbf{Appendix C}. 


 \begin{figure}[tb!]
\begin{center}
\includegraphics[width=\linewidth]{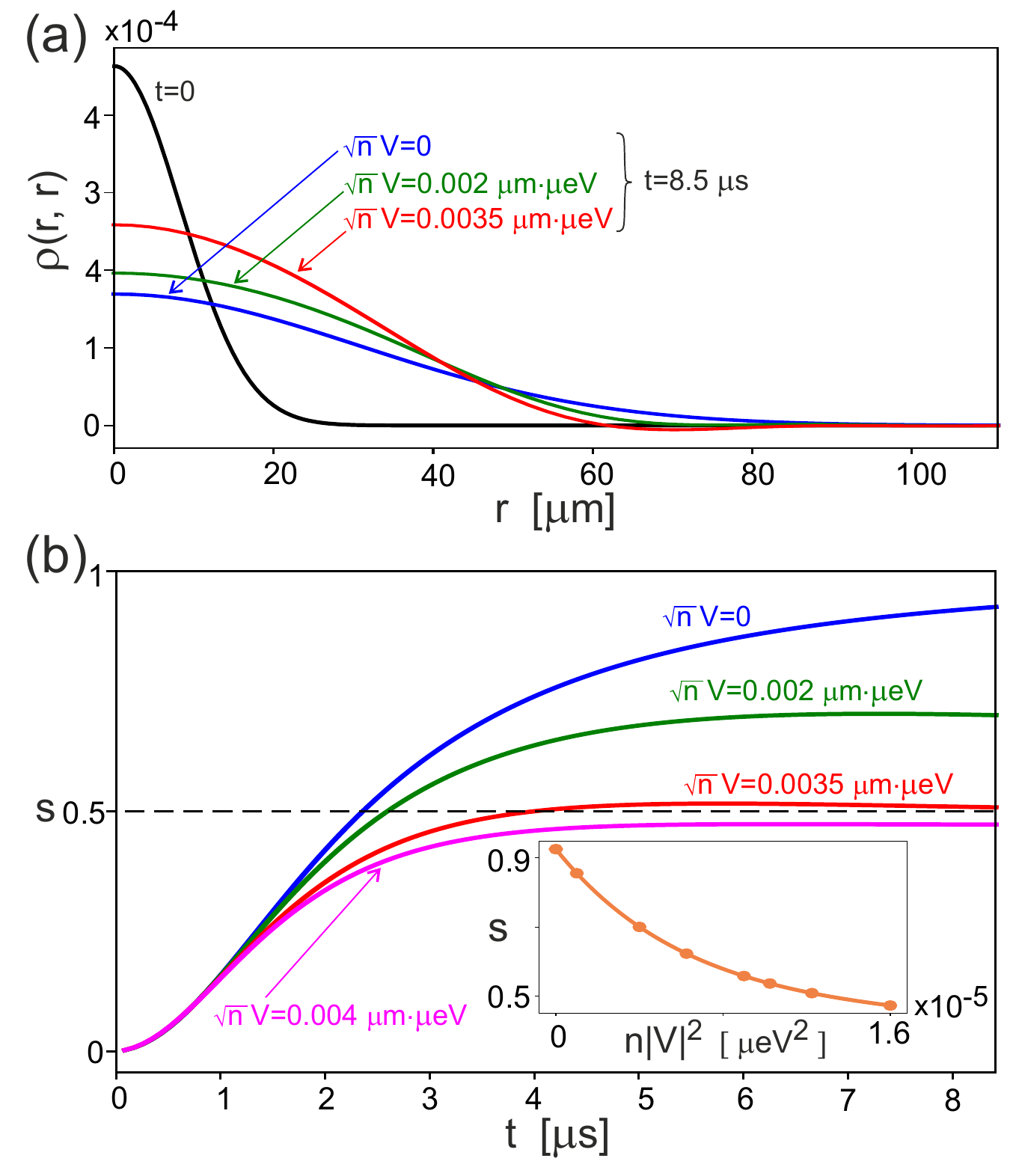}
\end{center}
\caption{ \label{pallet 2}
(a) Time evolution of an initially coherent excitonic wavepacket. Black solid line corresponds to the initial density distribution divided by factor $5$ to make the scale of the curve comparable to typical scale of the other dependencies, color lines correspond to the density profiles after $t =$ 8.5 \micro s for different disordered potential magnitudes $\sqrt{n} U$.  (b) The propagation exponent $s(t)$ defined by Eq.\ref{PropagationExponent} as function of time calculated for different disordered potential magnitudes $\sqrt{n} U$. The asymptotic value of this parameter close to $1$ characteristic for weak disorder indicates that the transport is ballistic. On the contrary, the asymptotic value close to $1/2$ characteristic for strong disorder is the signature of the purely diffusive propagation. Intermediate values correspond to mixed transport regime. The inset shows the dependence of the asymptotic value of the propagation exponent as function of the disorder strength. 
The quadratic dispersion of excitons were considered with the effective mass $m_x = 2m_e$.
}
\end{figure}

The results are shown in Fig. \ref{pallet 2}. Panel (a) illustrates the time evolution of an initially coherent wavepacket as function of the disorder strength $\sqrt{n}V$. One sees, that the increase of disorder slowers down the propagation as expected. 

An important parameter characterizing the propagation regime is the propagation exponent, calculated as 
\begin{equation}
s(t) = \frac{d\ln(<r>)}{d\ln (t)} \label{PropagationExponent}.
\end{equation}

For the ballistic propagation $s=1$ (see Eq.\ref{DeltaBallistic}), while for the diffusive $s=1/2$ (see Eq.\ref{DeltaDiffusive}). The dynamics of the propagation exponent is illustrated by the panel (b). One can clearly see that the increase of disorder leads to the gradual decrease of the asymptotic value of $s$ which corresponds to the crossover between ballistic and diffusive regimes. 

\section{Lower polariton transport}

Let us now introduce the coupling between the excitonic and photonic modes, setting $\Omega_R\neq0$. As it was mentioned, in this case upper and lower polariton branches $E_\pm(\bm{k})$ separated in energy by $\hbar\Omega_R$ appear (see Fig.1). Naturally, the transport on both of these branches strongly depends on corresponding photonic fraction and is thus defined by corresponding Hopfield coefficients. 

Let us notice that equation (\ref{Final Polaritons's Master equation}) contains both intra-band correlators $\rho^{++}(\bm{k}, \bm{k'}), \; \rho^{--}(\bm{k}, \bm{k'})$ defining the distributions of upper and lower polaritons in the real space and cross-band correlators  $\rho^{\pm, \mp}(\bm{k}, \bm{k}')$. We state the problem as an initial value problem with both intra-band correlators corresponding to the upper polariton branch and cross-band correlators equal to zero at $t = 0$. This allows to eliminate cross band correlations adiabatically \cite{brion2007adiabatic} in the limit of $\Omega_R \tau_0\gg 1$, where
\begin{equation}
\frac{1}{\tau_0} = \frac{2\pi}{\hbar} \frac{n}{A}|U|^2 \sum_{\bm{k}'}\delta(E_-(\bm{k'}) - E_-(\bm{k})).
\end{equation}

Indeed, putting 
\begin{equation}
\frac{\partial}{\partial t}\rho^{+-} = 0.  
\end{equation} 
we get
\begin{equation}
    \rho^{+-} \sim  \frac{S(\rho^{+-}, \rho^{++}, \rho^{--})}{(E_{+}(\bm{k}) - E_{-}(\bm{k'}))} \sim \rho^{--}\frac{|U^2|}{\Omega_R}
\end{equation}
Due to the fact that $(E_{+}(\bm{k}) - E_{-}(\bm{k'})) \sim \hbar\Omega_R$ and the scattering terms are bounded from above by $\tau_0^{-1}$  we can estimate
\begin{equation}
    \rho^{+-}  \sim \frac{\rho^{--}}{\tau_0\Omega_R},
\end{equation}
which vanishes in the limit if $\Omega_R \tau_0 \gg 1$, which is usually satisfied in realistic systems with moderate values of the disorder.  Also note, that if we assume the scattering to be purely elastic, for moderate disorders the inter-band scattering becomes impossible because the energy ranges of the upper and the lower polaritons do not overlap. 

Under these assumptions and in the s-scattering limit the expression for the scattering term $S$ in equation (\ref{Final Polaritons's Master equation}) reads for Upper and lower polaritons
\begin{eqnarray}
     && S^{\pm} = \pi\frac{n}{A}\hbar|U|^2 \rho^{\pm}(\bm{k}, \bm{k'}, t) \sum_{\bm{q}} \Big(\alpha_{\pm}^2(\bm{k})\alpha_{\pm}^2(\bm{k+q})\times \nonumber \\
     &&\times\delta(E(\bm{k})-E(\bm{k+q})) +  \nonumber \\
     && +\alpha_{\pm}^2(\bm{k'})\alpha_{\pm}^2(\bm{k'+q}) \delta(E(\bm{k'})-E(\bm{k'+q}))\Big )-  \\ 
     && -\pi\frac{n}{A}\hbar|U|^2\sum_{\bm{q}} \rho^{\pm}(\bm{k+q}, \bm{k'+q}, t)\alpha_{\pm}(\bm{k})\alpha_{\pm}(\bm{k+q})\times \nonumber \\
     && \times\alpha_{\pm}(\bm{k'})\alpha_{\pm}(\bm{k'+q})\Big( \delta(E(\bm{k})-E(\bm{k+q})) + \nonumber \\ 
     &&+\delta(E(\bm{k'})-E(\bm{k'+q})) \Big) \nonumber
     \label{polaritons scattering term}
\end{eqnarray}

The problem is thus reduced to the problem of the transport of scalar bosons with non-parabolic dispersion. In this work the focus is on lower polaritons, but the case of upper polaritons is fully analogical. As expected, the scattering integral contain the values of the Hopfield coefficients $\alpha_\pm(\bm{k})$, defined by the detuning between the excitonic and photonic modes $\delta$ (see Fig. 1). It worth noticing that equation (\ref{Final Polaritons's Master equation}) with scattering terms (\ref{scattering terms}) could be used for studying the transport in both strong and weak coupling regimes. However the cases of weak and intermediate coupling are beyond the scope of this paper.

We performed the numerical simulations of the dynamics of lower polaritons for the same initial conditions as in the previous section, creating initially a coherent excitonic Gaussian wavepacket given by Eq.\ref{InitialRho}. We focused on the dependence of the transport regime on detuning $\delta$, varying it in the interval in the interval [-50, -5] meV and set the other parameters as in the paper \cite{wurdack2021motional} ($m_{ph} = 0.8\cdot 10^{-5}m_e$ , $\hbar \Omega_R$ = 12.5 meV).

\begin{figure}[tb!]
\begin{center}
\includegraphics[width=\linewidth]{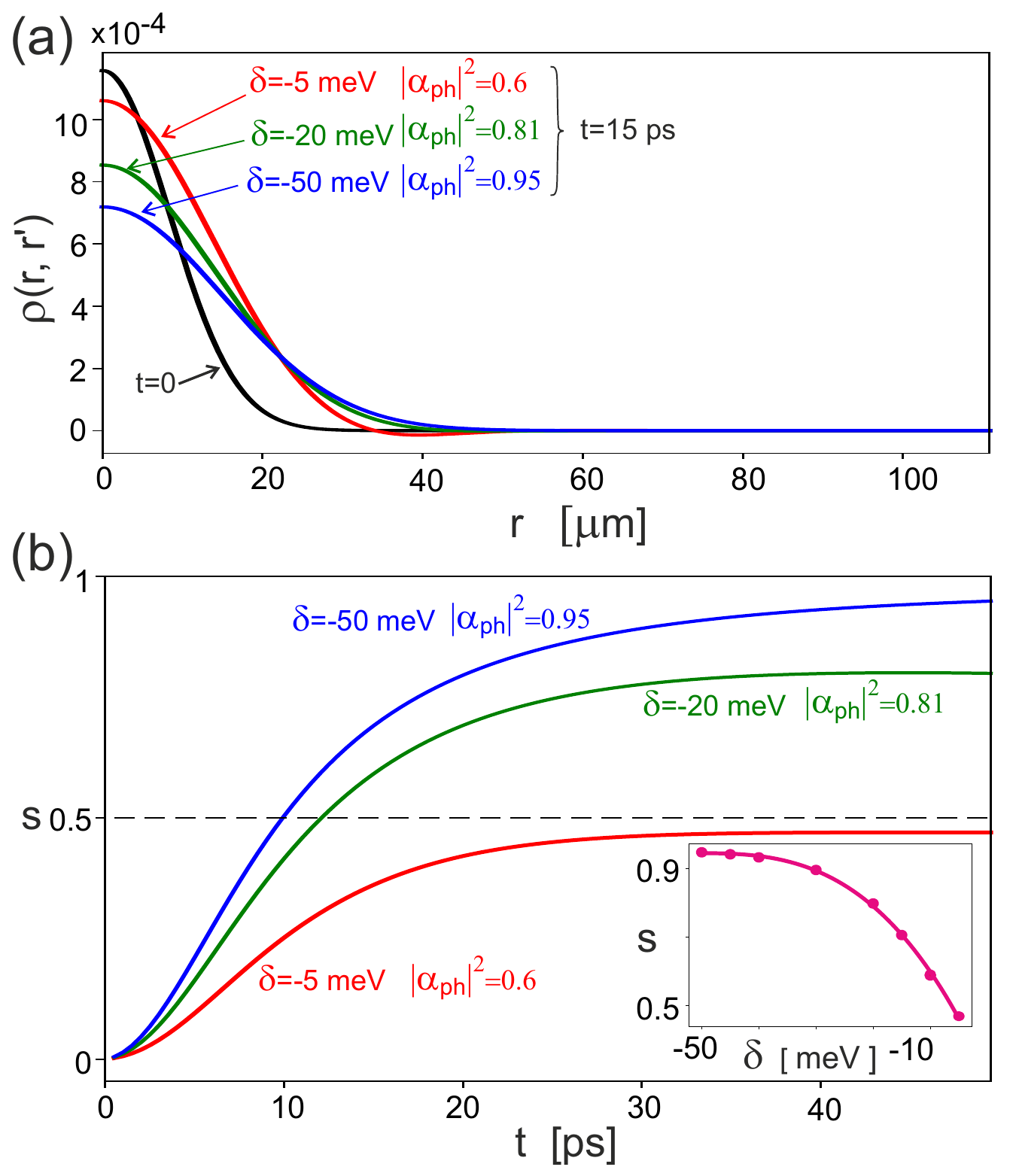}
\end{center}
\caption{ \label{pallet 3}
(a) Time evolution of an initially coherent polaritonic wavepacket. The black solid line corresponds to the initial density distribution divided by factor $2$ to make the scale of the curve comparable to the characteristics scales of the other curves, the blue, green and red lines correspond to the density profiles after $t =$ 15 ps for different values of the detuning $\delta$. The magnitude of the disorder potential is set to $\sqrt{n} U=1.5$ \micro m$\cdot$meV.  (b) The propagation exponent $s(t)$ defined by Eq.\ref{PropagationExponent} as function of time calculated for different detunings $\delta$. The asymptotic value of this parameter
close to 1 characteristic for big negative detunings and photonic character of lower polaritons indicates that the
transport is ballistic. On the contrary, the asymptotic value
close to 1/2 characteristic small negative detunings and the presence of a substantial excitonic fraction in lower polaritons corresponds to the diffusive propagation. Intermediate values
correspond to mixed transport regime. The inset shows the
dependence of the asymptotic value of the propagation expo-
nent as function of the detuning and is characteristic to the crossover from the ballistic to diffusive propagation in good agreement with resent experiments \cite{balasubrahmaniyam2023enhanced, wurdack2021motional}.}.
\end{figure}

The results are shown in Fig. \ref{pallet 3}. Panel (a) illustrates the time evolution of an initially coherent lower polariton wavepacket as function of the detuning $\delta$ for the fixed value of the strength $\sqrt{n}U=1.5$ \micro m$\cdot$meV. As one can see, the increase of the negative detuning leading to the increase of the photonic fraction in a lower polariton enhances the propagation. This is expected, as disorder scattering is relevant for the excitonic fraction only. 

The dynamics of the propagation exponent for lower polaritons is illustrated by the panel (b). One can clearly see that the increase of negative detuning $\delta$ increases the asymptotic value of $s$ and thus corresponds to the crossover between diffusive and ballistic regimes, as it was recently reported experimentally \cite{balasubrahmaniyam2023enhanced, wurdack2021motional}.

We can also derive semi-classical kinetic equations for the upper and lower polaritons, introducing the semi quasi-classical probability distributions with the help of Wigner representation \cite{altland2010condensed} which could be applied under condition $\frac{<E>\tau_0}{\hbar} >> 1$ at timescales comparable to the relaxation time.
\begin{equation}
\rho_\pm(\bm{k}, \bm{r}, t) = \int d^2\bm{\kappa} \rho_\pm(\bm{k} + \frac{\kappa}{2}, \bm{k} - \frac{\kappa}{2}, t)\exp(i\bm{\kappa}\bm{r})
\end{equation}

Assuming the scattering to be elastic so that transitions between the upper and the lower bands are forbidden, and adiabatically eliminating cross-band correlations one gets from Eqs.\ref{Final Polaritons's Master equation} and \ref{scattering terms}:
\begin{equation}
        \frac{\partial}{\partial_t}\rho_\pm(\bm{k}, \bm{r}, t) + \bm{v_\pm(k)}\cdot\bm{\nabla}\rho_\pm(\bm{k}, \bm{r}, t) = S^\pm(\rho_\pm(\bm{k}, \bm{r}, t)),
        \label{kinetic equation}
\end{equation}
where the scattering integrals are 
\begin{eqnarray}
    &&S^{\pm} = -\sum_{\bm{q}} \frac{2\pi}{\hbar}\frac{n}{A}\alpha^2_{\pm}(\bm{k+q})\alpha^2_{\pm}(\bm{k}) |U_{\bm{q}}|^2\times \\ 
    &&\times\delta(E_{\pm}(\bm{k+q})-E_{\pm}(\bm{k}))(\rho_\pm(\bm{k}, \bm{r}, t) - \rho_\pm(\bm{k+q}, \bm{r}, t)) \notag
\end{eqnarray}

These kinetic equations are very similar to the kinetic equations for excitons. The main difference is the presence of  Hopfield coefficients in the scattering amplitudes and the difference of the effective masses of the particles (excitons and polaritons) by several orders of magnitude. Let us remark that kinetic equation (\ref{kinetic equation}) describes both ballistic and diffusive transport regimes, however, it doesn’t account for quantum effects such as interference.

In the considered limit one can introduce the relaxation times \cite{lifschitz1983physical} for excitons and polaritons as: 
\begin{equation}
    \frac{1}{\tau_0(\bm{k})} = \sum_{\bm{k'}}W_{\bm{k'k}},
    \label{relaxation time definition}
\end{equation}
where scattering rates for the excitons and polaritons are given by :
\begin{equation}
    W^{x}_{\bm{k'k}} = \frac{2\pi}{\hbar} \frac{n}{A}|U|^2 \delta(\varepsilon_x(\bm{k'}) - \varepsilon_x(\bm{k})),
\end{equation}
\begin{equation}
    W^{p}_{\bm{k'k}} = \frac{2\pi}{\hbar} \frac{n}{A}|U|^2 \alpha^2(\bm{k'})\alpha^2(\bm{k})  \delta(E_p(\bm{k'}) - E_p(\bm{k}))
\end{equation}

It can be easily seen that the ratio of the excitonic and polaritonic relaxation times is 
\begin{equation}
    \frac{\tau_{x}(\bm{k})}{\tau^{\pm}_p(\bm{k})} = \alpha_{\pm}^4(\bm{k})\frac{D_p(\bm{k})}{D_{x}(\bm{k})} = \alpha_{\pm}^4(\bm{k})\frac{m_p(\bm{k})}{m_{x}(\bm{k})},
\end{equation}
where $D_x(\bm{k}), D_p(\bm{k})$ - excitonic and polaritonic densities of states.

\begin{figure}[tb!]
\begin{center}
\includegraphics[width=\linewidth]{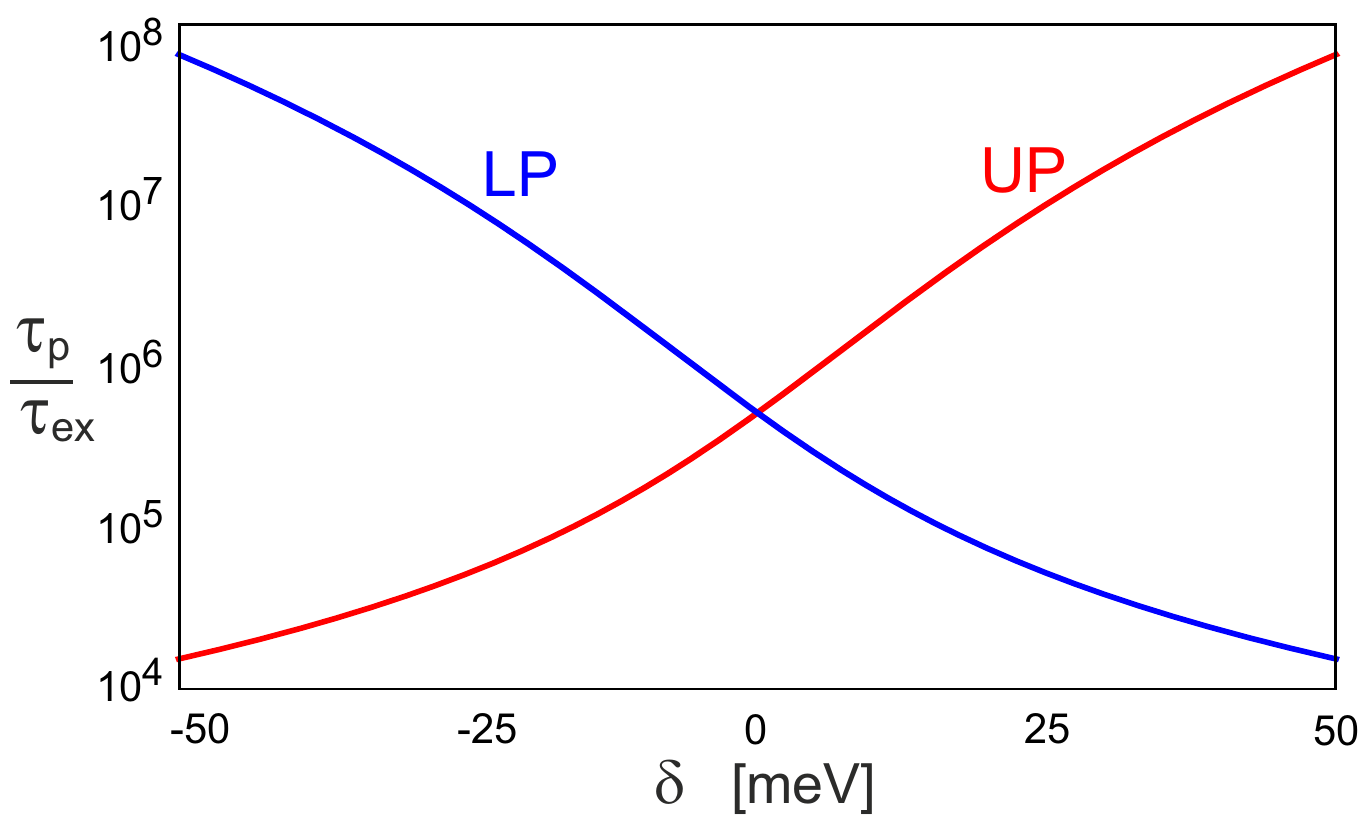}
\end{center}
\caption{ \label{relaxation times}
The ratios of polaritonic and excitonic relaxation times as function of the detuning $\delta$. The blue line corresponds to the lower and the red line to the upper polariton branches. 
It is seen that the ratio varies by four order of magnitude if the detuning $\Delta E$ changes from $-50$ meV to $50$ meV. Such variation of the relaxation times is caused by the change of the effective mass and Hopfield coefficients for upper and lower polaritons. 
}
\end{figure}

The dependencies of the relaxation times of lower and upper polaritons on detuning $\delta$ are shown in Fig.~\ref{relaxation times}. As expected, polaritonic relaxation times exceed the excitonic relaxation times by several orders of magnitude for realistic experimental conditions. Naturally, the increase of photonic fraction leads to the decrease of the relaxation time. Also we would like to note that  the kinetic equations (\ref{kinetic equation}) could be used to study weak localization of polariotons \cite{dyakonov1994magnetoconductance, dmitriev1997nonbackscattering}.

\section{Conclusion}
In conclusion, we considered the dynamics of polaritons in a planar microcavity with short range excitonic disorder. Basing on the master equation for full density matrix of the system and treating the disorder scattering in Born-Markov approximation we analyzed the crossover between ballistic and diffusive regimes of the polariton transport with change of the strength of the scattering potential and the detuning between excitonic and photonic modes. We also demonstrated that semiclassical kinetic equations and relaxation times can be obtained for polaritons. Our results are in good agreement with experimental data reported in Refs. \cite{balasubrahmaniyam2023enhanced, wurdack2021motional}.

\section{Acknowledgements}
This work was supported by the Ministry of Science and Higher Education of Russian Federation, goszadanie no. 2019-1246 and Priority 2030 Federal Academic Leadership Program. AVY and IAS thank Icelandic Research Fund (Rannis) for the support in frameworks of the project No. 163082-051.

\begin{widetext}

    \section{Appendix A: Taking average on random impuritie's positions}
We consider uncorrelated delta-functional impurities 
\begin{equation}
    V_{\bm{kk'}} = \frac{U_{\bm{kk'}}}{A}\sum_i(\exp(-i(\bm{k}-\bm{k'})\bm{R}_i))
    \label{noncorreleted impurities}
\end{equation}
where $U_{\bm{kk'}} = \int d^2\bm{r} e^{i\bm{k-k'}{r}}V(r)$ - potential Fourier component, $A$- sample area. Factor $1/A$ in the above expression appear due to the plane wave normalization $\psi_{k} = \frac{e^{i\bm{k}\bm{r}}}{\sqrt{A}}$ in the expression for matrix component $V_{\bm{kk'}} = \bra{\psi_{\bm{k}}}\hat{V}\ket{\psi_{\bm{k'}}}$. For this dependency of $V_{kk'} $ on $k$, $k'$ the averaging can be done analytically representing the averaged term as  
\begin{equation}
    \langle\sum V_{\bm{k''k'''}}V_{\bm{kk'}}[c_{\bm{k''}}^{\dagger} c_{\bm{k'''}}, [c_{\bm{k}}^{\dagger} c_{\bm{k'}}, \rho]]\rangle_c = \sum \left <\sum_{i, j} (\exp(-i(\bm{k}-\bm{k'})\bm{R}_i)(\exp(-i(\bm{k''}-\bm{k'''})\bm{R}_j) \right>_c \frac{U_{\bm{k''k'''}}U_{\bm{kk'}}}{A^2}[c_{\bm{k''}}^{\dagger} c_{\bm{k'''}}, [c_{\bm{k}}^{\dagger} c_{\bm{k'}}, \rho]].
    \label{quadratic terms averaging}
\end{equation}
In the latter expression the inner sum can be calculated 
\begin{equation}
   \left <\sum_{i, j} (\exp(-i(\bm{k}-\bm{k'})\bm{R}_i)(\exp(-i(\bm{k''}-\bm{k'''})\bm{R}_j) \right>_c =  N_i \delta_{\bm{k}-\bm{k'} + \bm{k''} - \bm{k'''}, 0} = N_i\delta_{\bm{q}_1 + \bm{q}_2, 0},
   \label{chaos correlations}
\end{equation}
where $N_i$ - impurity's quantity. 
Then, finally, we obtain the expression for the averaged term 
\begin{equation}
    \left<\sum V_{\bm{k''k'''}}V_{\bm{kk'}}[c_{\bm{k''}}^\dagger c_{\bm{k'''}}, [c_{\bm{k}}^\dagger c_{\bm{k'}}, \rho]]\right>_c = \frac{n}{A}\sum_{\bm{k}, \bm{k'}, \bm{q}}U_{\bm{k'},\bm{k'-q}}U_{\bm{k},\bm{k+q}}[c_{\bm{k'}}^\dagger c_{\bm{k'-q}}, [c_{\bm{k}}^\dagger c_{\bm{k+q}}, \rho]]
    \label{chaos average}
\end{equation}
where $n = \frac{N_i}{A}$ - impurities concentration.

\section{Appendix B: Detailed derivation of Master equation for the polaritons}
First the expression for $H^I_V(t)$ have to be derived. Do to this we can express the $H^I_V(t)$ using the exponent of adjoint representation of $H_0$
\begin{equation}
    H^I_V(t) = \exp(\frac{i}{\hbar} H_0 t)H_V \exp(-\frac{i}{\hbar} H_0 t) = \exp(\frac{i}{\hbar} t\cdot \bm{ad_{H_0}})H_V,
\end{equation}
than we calculated corresponding correlators
\begin{equation}
    [H_0, {c_{\bm{k}}}_{\sigma_1}^{\dagger}{c_{\bm{k'}}}_{\sigma_2}] = (E_{\sigma_1}(\bm{k})-E_{\sigma_2}(\bm{k'})) {c_{\bm{k}}}_{\sigma_1}^{\dagger}{c_{\bm{k'}}}_{\sigma_2},
\end{equation}
from this expression it is obvious that we will have a tailor series for the $e^{E_{\sigma_1}(k)-E_{\sigma_2}(k')}$. Finally using formula \ref{chaos average} we can calculate the integral and obtain the expression for the scattering part of the Master equation  (\ref{general Master equation})
\begin{eqnarray}
     &&\langle M_0[H^I_V(t), \int^t_0 dt'[H^I_V(t'), \rho^I(t)]]M_0^\dagger\rangle_c = 
     \frac{n}{A}\sum_{\bm{k}, \bm{k'}, \bm{q}}U_{\bm{k'},\bm{k'-q}}U_{\bm{k},\bm{k+q}}  \\
     &&\sum_{\sigma_1, \sigma_2, s1, s2 = \pm 1} \sigma_1 \cdot \sigma_2\cdot s_1 \cdot s_2 \alpha_{s_1}(\bm{k'})\alpha_{s_2}(\bm{k'-q})\alpha_{\sigma_1}(\bm{k})\alpha_{\sigma_2}(\bm{k+q})f_{\bm{k}, \bm{k+q}}^{\sigma_1, \sigma_2}(t)[{c_{\bm{k'}}}_{s_1}^{\dagger}{c_{\bm{k'-q}}}_{s_2}, [{c_{\bm{k}}}_{\sigma_1}^{\dagger}{c_{\bm{k+q}}}_{\sigma_2}, \rho]], \nonumber
     \label{H_V^I}
\end{eqnarray}
where $f(t)$ is given by (\ref{fDefinition}).

The dynamics of the polaritonic pulses in the system can be studied by writing evolution equations for correlators $\rho_{\zeta_1, \zeta_2}(k, k', t) = \langle{c^\dagger_{\bm{k}}}_\pm {c_{\bm{k'}}}_\pm\rangle = \text{Tr}(\rho {c^\dagger_k}_{\zeta_1} {c_{k'}}_{\zeta_2})$. To do this we use the following fact
\begin{eqnarray}
    &&\text{Tr}([{c_{\bm{k'}}}_{s_1}^{\dagger}{c_{\bm{k'-q}}}_{s_2}, [{c_{\bm{k}}}_{\sigma_1}^{\dagger}{c_{\bm{k+q}}}_{\sigma_2}, \rho]]F) =  \notag \\
    &&=\text{Tr}(\rho[{c_{\bm{k}}}_{\sigma_1}^{\dagger}{c_{\bm{k+q}}}_{\sigma_2}, [{c_{\bm{k'}}}_{s_1}^{\dagger}{c_{\bm{k'-q}}}_{s_2}, F]]),
    \label{operator averaging}
\end{eqnarray}
where $F$ - arbitrary operator. To proof this expression we need to use several times cycle permutations under the trace operation
\begin{eqnarray}
    &&\text{Tr}([{c_{\bm{k'}}}_{s_1}^{\dagger}{c_{\bm{k'-q}}}_{s_2}, [{c_{k}}_{\sigma_1}^{\dagger}{c_{\bm{k+q}}}_{\sigma_2}, \rho]]F) = \text{Tr}([{c_{\bm{k'}}}_{s_1}^{\dagger}{c_{
    \bm{k'-q}}}_{s_2}[{c_{\bm{k}}}_{\sigma_1}^{\dagger}{c_{\bm{k+q}}}_{\sigma_2}\rho - \rho{c_{\bm{k}}}_{\sigma_1}^{\dagger}{c_{\bm{k+q}}}_{\sigma_2}] F) = \notag\\ \notag 
    && \text{Tr}\left[\left({c_{\bm{k'}}}_{s_1}^{\dagger}{c_{\bm{k'-q}}}_{s_2}{c_{\bm{k}}}_{\sigma_1}^{\dagger}{c_{\bm{k+q}}}_{\sigma_2}\rho + \rho{c_{\bm{k}}}_{\sigma_1}^{\dagger}{c_{\bm{k+q}}}_{\sigma_2}{c_{\bm{k'}}}_{s_1}^{\dagger}{c_{\bm{k'-q}}}_{s_2} - {c_{\bm{k'}}}_{s_1}^{\dagger}{c_{\bm{k'-q}}}_{s_2}\rho{c_{\bm{k}}}_{\sigma_1}^{\dagger}{c_{\bm{k+q}}}_{\sigma_2} - {c_{\bm{k}}}_{\sigma_1}^{\dagger}{c_{\bm{k+q}}}_{\sigma_2}\rho{c_{\bm{k'}}}_{s_1}^{\dagger}{c_{\bm{k'-q}}}_{s_2} \right)F\right] = \\ \notag
    &&\text{Tr}\left[\rho \left(F{c_{\bm{k'}}}_{s_1}^{\dagger}{c_{\bm{k'-q}}}_{s_2}{c_{\bm{k}}}_{\sigma_1}^{\dagger}{c_{\bm{k+q}}}_{\sigma_2}+ {c_{\bm{k}}}_{\sigma_1}^{\dagger}{c_{\bm{k+q}}}_{\sigma_2}{c_{\bm{k'}}}_{s_1}^{\dagger}{c_{\bm{k'-q}}}_{s_2}F - {c_{\bm{k}}}_{\sigma_1}^{\dagger}{c_{\bm{k+q}}}_{\sigma_2}F{c_{\bm{k'}}}_{s_1}^{\dagger}{c_{\bm{k'-q}}}_{s_2} - {c_{\bm{k'}}}_{s_1}^{\dagger}{c_{\bm{k'-q}}}_{s_2}F{c_{\bm{k}}}_{\sigma_1}^{\dagger}{c_{\bm{k+q}}}_{\sigma_2} \right)\right] = \\ 
    &&Tr(\rho[{c_{\bm{k}}}_{\sigma_1}^{\dagger}{c_{\bm{k}}}_{\sigma_2}, [{c_{\bm{k'}}}_{s_1}^{\dagger}{c_{\bm{k'-q}}}_{s_2}, F]])
    \label{proof of operator average}
\end{eqnarray}
Finally, substituting ${c_{\bm{p}}}_{\zeta_1}^{\dagger}{c_{\bm{p'}}}_{\zeta_2}$ operator instead $F$ and taking into account correlation relations we get the following expression for correlators (\ref{correlators}) which contains four terms corresponding to $S_{1-4}$ 
\begin{eqnarray}
    &&[{c_{\bm{k}}}_{\sigma_1}^{\dagger}{c_{\bm{k+q}}}_{\sigma_2}, [{c_{\bm{k'}}}_{s_1}^{\dagger}{c_{\bm{k'-q}}}_{s_2}, {c_{\bm{p}}}_{\zeta_1}^{\dagger}{c_{\bm{p'}}}_{\zeta_2}]] = [{c_{\bm{k}}}_{\sigma_1}^{\dagger}{c_{\bm{k+q}}}_{\sigma_2}, ({c_{\bm{k'}}}_{s_1}^{\dagger}[{c_{\bm{k'-q}}}_{s_2}, {c_{\bm{p}}}_{\zeta_1}^{\dagger}]{c_{\bm{p'}}}_{\zeta_2} + {c_{\bm{p}}}_{\zeta_1}^{\dagger}[{c_{\bm{k'}}}_{s_1}^{\dagger}, {c_{\bm{p'}}}_{\zeta_2}]{c_{\bm{k'-q}}}_{s_2})] = \notag\\ \notag
    &&[{c_{\bm{k}}}_{\sigma_1}^{\dagger}{c_{\bm{k+q}}}_{\sigma_2}, \left({c_{\bm{k'}}}_{s_1}^{\dagger}{c_{\bm{p'}}}_{\zeta_2}\delta_{s_2, \zeta_1}\delta_{\bm{p} -(\bm{k'}-\bm{q}), 0} - {c_{\bm{p}}}_{\zeta_1}^{\dagger}{c_{\bm{k'-q}}}_{s_2}\delta_{s_1, \zeta_2}\delta_{\bm{p'} - \bm{k'}, 0}\right)] = \\ \notag
    &&\delta_{s_2, \zeta_1}\delta_{\bm{p} -(\bm{k'-q}), 0}({c_{\bm{k}}}_{\sigma_1}^{\dagger}[{c_{\bm{k+q}}}_{\sigma_2},  {c_{\bm{k'}}}_{s_1}^{\dagger}]{c_{\bm{p'}}}_{\zeta_2} + 
    {c_{\bm{k'}}}_{s_1}^{\dagger}[{c_{\bm{k}}}_{\sigma_1}^{\dagger}, {c_{\bm{p'}}}_{\zeta_2}]{c_{\bm{k+q}}}_{\sigma_2}) - \\ 
    &&\delta_{s_1, \zeta_2}\delta_{\bm{p'} - \bm{k'}, 0} ({c_{\bm{k}}}_{\sigma_1}^{\dagger}[{c_{\bm{k+q}}}_{\sigma_2}, {c_{\bm{p}}}_{\zeta_1}^{\dagger}]{c_{\bm{k'-q}}}_{s_2} + 
    {c_{\bm{p}}}_{\zeta_1}^{\dagger}[{c_{\bm{k}}}_{\sigma_1}^{\dagger}, {c_{\bm{k'-q}}}_{s_2}]{c_{\bm{k+q}}}_{\sigma_2}) = \\ \notag
     &&\delta_{s_2, \zeta_1}\delta_{\bm{p} -(\bm{k'}-\bm{q}), 0}({c_{\bm{k}}}_{\sigma_1}^{\dagger}{c_{\bm{p'}}}_{\zeta_2}\delta_{s_1, \sigma_2}\delta_{\bm{k'} -(\bm{k}+\bm{q}), 0} - 
      {c_{\bm{k'}}}_{s_1}^{\dagger}{c_{\bm{k+q}}}_{\sigma_2}\delta_{\zeta_2, \sigma_1}\delta_{\bm{p'} -\bm{k}, 0}) - \\ \notag
      && \delta_{s_1, \zeta_2}\delta_{\bm{p'} - \bm{k'}, 0}({c_{\bm{k}}}_{\sigma_1}^{\dagger}{c_{\bm{k'-q}}}_{s_2} \delta_{\sigma_2, \zeta_1}\delta_{\bm{p} - (\bm{k}+\bm{q}), 0}) - 
      {c_{\bm{p}}}_{\zeta_1}^{\dagger} {c_{\bm{k+q}}}_{\sigma_2}\delta_{\sigma_1, s_2}\delta_{\bm{k} - (\bm{k'}-\bm{q}),0})
      \label{correlators}
\end{eqnarray}

Substituting the expression (\ref{correlators}) to the (\ref{H_V^I}) we obtained the final result (\ref{Final Polaritons's Master equation}) with $S = S_1+S_2+S_3+S_4$ were $S_{1-4}$ are given by the following expressions

\begin{subequations}
    \begin{eqnarray}
        &&S_1 = \frac{n}{A}\sum_{\bm{q}}U_{\bm{k+q},\bm{k}}U_{\bm{k},\bm{k+q}} \sum_{\sigma_1, \sigma_2 = \pm} \sigma_1  \cdot \zeta_1 \alpha_{\sigma_2}(k+q)\alpha_{\zeta_1}(k)\alpha_{\sigma_1}(\bm{k})\alpha_{\sigma_2}(\bm{k+q})f_{\bm{k}, \bm{k+q}}^{\sigma_1, \sigma_2}(t)\rho_{\sigma_1, \zeta_2}(\bm{k}, \bm{k'}, t)  \\
        &&S_2 = \frac{n}{A}\sum_{\bm{q}}U_{\bm{k'},\bm{k'-q}}U_{\bm{k'-q},\bm{k'}} \sum_{\sigma_1, \sigma_2 = \pm} \sigma_2  \cdot \zeta_2 \alpha_{\zeta_2}(\bm{k'})\alpha_{\sigma_1}(\bm{k'-q})\alpha_{\sigma_1}(\bm{k'-q})\alpha_{\sigma_2}(\bm{k'})f_{\bm{k'-q}, \bm{k'}}^{\sigma_1, \sigma_2}(t) \times\\
        &&\times\rho_{\zeta_1, \sigma_2}(\bm{k}, \bm{k'}, t) \nonumber \\
        &&S_3 = \frac{n}{A}\sum_{\bm{q}}U_{\bm{k+q},\bm{k}}U_{\bm{k'},\bm{k'+q}} \sum_{\sigma_1, \sigma_2 = \pm} \sigma_1 \cdot \sigma_2  \cdot \zeta_1\cdot \zeta_2 \alpha_{\sigma_1}(\bm{k+q})\alpha_{\zeta_1}(\bm{k})\alpha_{\zeta_2}(\bm{k'})\alpha_{\sigma_2}(\bm{k'+q})f_{\bm{k'}, \bm{k'+q}}^{\zeta_2, \sigma_2}(t) \times \\
        &&\times\rho_{\sigma_1, \sigma_2}(\bm{k+q}, \bm{k'+q}, t) \nonumber \\        
        &&S_4 = \frac{n}{A}\sum_{\bm{q}}U_{\bm{k'},\bm{k'-q}}U_{\bm{k-q},\bm{k}} \sum_{\sigma_1, \sigma_2 = \pm} \sigma_1 \cdot \sigma_2  \cdot \zeta_1\cdot \zeta_2 \alpha_{\zeta_2}(\bm{k'})\alpha_{\sigma_2}(\bm{k'-q})\alpha_{\sigma_1}(\bm{k-q})\alpha_{\zeta_1}(k)f_{\bm{k-q}, \bm{k}}^{\sigma_1, \zeta_1}(t)\times \\
        &&\times\rho_{\sigma_1, \sigma_2}(\bm{k-q}, \bm{k'-q}, t) \nonumber.
    \end{eqnarray}
    \label{scattering terms}
\end{subequations}

The functions 
\begin{equation}
f^{\sigma_1, \sigma_2}_{k, k'}(t) =\frac{1-\exp(-\frac{i}{\hbar}t(E_{\sigma_1}(k)-E_{\sigma_2}(k'))}{\frac{i}{\hbar}(E_{\sigma_1}(k)-E_{\sigma_2}(k'))}.  \label{fDefinition} 
\end{equation}
The particle transport can be described within the Born-Markov approximation in provided that $\frac{<E>\tau_0}{\hbar} >> 1$ where $<E>$ denotes average particles energy, $\tau_0$ is the relaxation time defined in (\ref{relaxation time definition}) \cite{lifschitz1983physical}. 
\begin{equation}
   f^{\sigma_1, \sigma_2}_{k, k'}(t) \xrightarrow{t>>\Delta_E}\pi \hbar \delta(E_{\sigma_1}(k)-E_{\sigma_2}(k')), \label{fAsymptotics}
\end{equation} 
and correspond therefore to the energy conservation during a scattering act.

\section{Appendix C: Numerical procedure}
\subsection{Numerical procedure for only excitonic problem}
In the 2D microcavity the elasticity of scattering fixes the length of the wave vector after scattering $|\tilde{\bm{k}}| = |\bm{k}|$. Thus we can move from integration with respect to $d^2 \bm{q}$ to integration with respect to $\bm{\tilde{k}} = \bm{k} + \bm{q}$. And then moving to the polar coordinates and considering properties of $\delta$ functions the following formula for (\ref{Excitonic scattering integral in elastic limit}) were obtained

\begin{equation}
    S_{2D}(\bm{k}, \bm{k'}) = \pi n \hbar |U|^2\rho(\bm{k}, \bm{k'}, t)(D_{2D}(\bm{k}) + D_{2D}(\bm{k'})) - \pi n \hbar |U|^2(I_{\bm{k}} + I_{\bm{k}'})
    \label{S_2D}
\end{equation}
\begin{equation}
    I_{\bm{k}}  = \frac{1}{(2\pi)^2} \int  d\theta_{\tilde{\bm{k}}}\left[k  \frac{ \rho(\tilde{\bm{k}}, \bm{k'}- \bm{k} + \tilde{\bm{k}}, t)}{\hbar|v_k|}\right],
\end{equation}
\begin{equation}
    I_{\bm{k'}} = \frac{1}{(2\pi)^2} \int d\theta_{\tilde{\bm{k}'}}\left[k' \frac{ \rho(\bm{k}- \bm{k'} + \tilde{\bm{k}}, \tilde{\bm{k}'}, t)}{\hbar|v_{k'}|}\right],
\end{equation} 

where $D_{2D}(\bm{k})$ is a density of states. The equation (\ref{Excitonic Master equation}) in this case is can be reduced to the system of ODEs with time independent coefficients by introducing a mesh in $\bm{k},\, \bm{k}'$ space. However, if we take $N$ points for $|\bm{k}|$ discretization and $N_{\theta}$ for $\theta$ discretization we have a total number of $N^2N_{\theta}^2$ parameters for density matrix $\rho(\bm{k}, \bm{k'})$ and $N^2N_{\theta}^3$ complexity of calculation of the ODE's system right part which makes this numerical problem enormous. Fortunately, for the case when initial reversal space distribution profile has axial symmetry such as Gaussian profile with $k_0 = 0$ it is possible to reduce total number parameters to the $N^2N_{\theta}$ and total right part calculation complexity to $N^2N_{\theta}^2$ which is much better. In the axial symmetric case the density matrix has symmetry $\rho(\hat{R}(\theta)\bm{k}, \hat{R}(\theta)\bm{k}', t) = \rho(\bm{k}, \bm{k}', t)$, so we can for example fix direction of $\bm{k} = [|\bm{k}|, 0]$ and reduce number of parameters. Finally, the equidistant mesh in $k$, $k'$, $\theta_{\bm{k'}}$ were considered. The integrals in (\ref{S_2D}) was approximately replaced by sums and interpolation were used to approximate points appear in the integrals approximations which do not contained in the mesh. The equation
\begin{eqnarray}  
    \rho(r, r, t) \sim \int dk dk' d\theta d\theta' k k' e^{-ik r\cos(\theta) + ik' r\cos(\theta' + \theta) }\rho(k, k', \theta', t) = \int dk dk' d\theta' k k' 2\pi J_0(q(k, k', \theta')r)\rho(k, k', \theta', t)
\end{eqnarray}
where $J_0$ - zero order Bessel function, $q =|\bm{k} - \bm{k'}| = \sqrt{{k'}^2 + k^2 - 2kk'\cos(\theta')} $ gives us the real space distribution. The accurate calculation of this integral with function oscillating very fast is also an ambitious task. The normalization condition $2\pi\int dx x \rho(x, x, 0) = 1$ were taken.

\subsection{Numerical procedure for lower band polaritons}
For the lower band polaritons the Master equation looks as follows
\begin{eqnarray}
     S = \pi \hbar \frac{n}{A}|U|^2 \sum_{\bm{q}} \left(\rho(\bm{k}, \bm{k'}, t)(\alpha^2(k)\alpha^2(\bm{k}+\bm{q}) \delta(E(\bm{k})-E(\bm{k}+\bm{q})) + \alpha^2(\bm{k'})\alpha^2(\bm{k'}+\bm{q}) \delta(E(\bm{k'})-E(\bm{k'}+\bm{q}))) \right) - \notag \\ 
     \pi \hbar \frac{n}{A}|U|^2 \sum_{\bm{q}} \rho(\bm{k}+\bm{q}, \bm{k'}+\bm{q}, t)\alpha(\bm{k})\alpha(\bm{k}+\bm{q})\alpha(\bm{k'})\alpha(\bm{k'}+\bm{q})(\delta(E(\bm{k})-E(\bm{k}+\bm{q})) + \delta(E(\bm{k'})-E(\bm{k'}+\bm{q}))),
\end{eqnarray}
Which result in equations are similar to (\ref{S_2D}) with inclusion of Hopfield coefficients $\alpha$. The numerical procedure stay the same.

 \begin{figure}[tb!]
\begin{center}
\includegraphics[width=\linewidth]{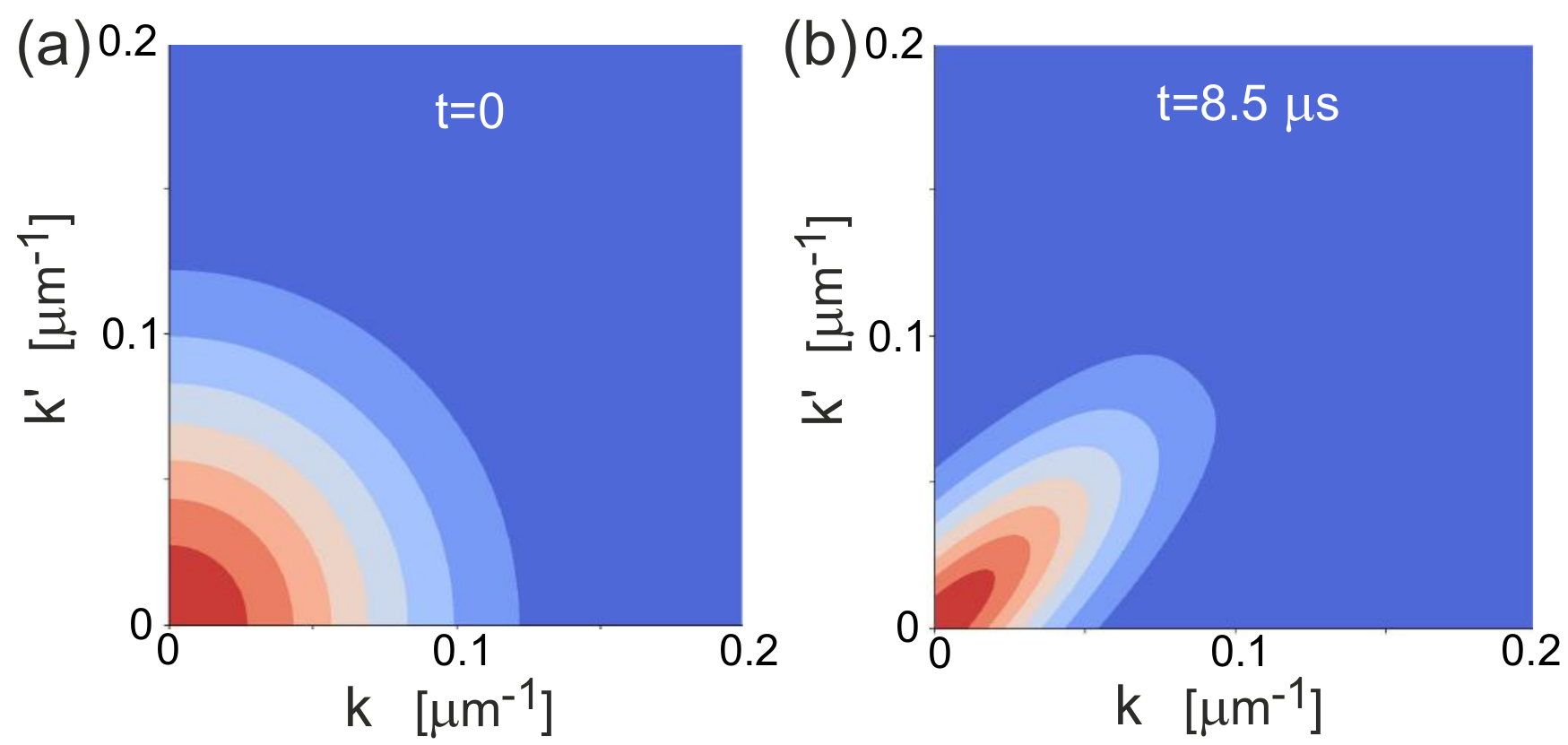}
\end{center}
\caption{ \label{k_r distributions}
The pallet shows the initial $\rho(k_r, k_r', \theta = 0, t=0)$ and final $\rho(k_r, k_r', \theta = 0, t=8.5\ \upmu \text{s})$ $k_r$ space distribution profiles for the diffusive case $\sqrt{n}U =$ 0.004 \micro eV $\cdot$ \micro m. In the diffusive case profile constricts to the line $k_r = k_r'$ which corresponds to field becoming more classical.
One can recall that kinetic equation for semiclassical probability distribution $\rho(k, r, t)$ can be derived from the Master equation (\ref{Excitonic Master equation}) using Wigner representation \cite{altland2010condensed} $\rho(k+\frac{\kappa}{2}, k - \frac{\kappa}{2}, t)$ and making expansion on small $\kappa$. 
}
\end{figure}

\end{widetext}

\bibliography{APFBibl}
 
\end{document}